\documentclass[twocolumn,secnumarabic,amssymb, amsmath, nobibnotes, aps, prx,superscriptaddress,showpacs,longbibliography]{revtex4-1}

\usepackage{graphicx}
\usepackage{footnote}
\usepackage{bm}%
\usepackage{color,soul}
\usepackage[colorlinks=true,linkcolor=blue]{hyperref}
\expandafter\ifx\csname package@font\endcsname\relax\else
 \expandafter\expandafter
 \expandafter\usepackage
 \expandafter\expandafter
 \expandafter{\csname package@font\endcsname}%
 \usepackage{dblfloatfix}%
\fi

\begin{document}

\title{Low-disorder microwave cavity lattices for quantum simulation with photons}%

\author{D.\ L.\ Underwood}
\affiliation{Department of Electrical Engineering, Princeton University, Princeton, New Jersey 08544}
\author{W.\ E.\ Shanks}
\affiliation{Department of Electrical Engineering, Princeton University, Princeton, New Jersey 08544}
\author{ Jens Koch}
\affiliation{Department of Physics and Astronomy, Northwestern University, Evanston, Illinois 60208}
\author{ A.\ A.\ Houck}
\affiliation{Department of Electrical Engineering, Princeton University, Princeton, New Jersey 08544}%

\date{\today}%
\begin{abstract}
We assess experimentally the suitability of coupled transmission line resonators for studies of quantum phase transitions of light. We have measured devices with low photon hopping rates $t/2\pi=0.8\,$MHz to quantify disorder in individual cavity frequencies. The observed disorder is consistent with small imperfections in fabrication.  We studied the dependence of the disorder on transmission line geometry and used our results to fabricate devices with disorder less than two parts in $10^4$. The normal mode spectrum of devices with a high photon hopping rate $t/2\pi=31\,$MHz shows little effect of disorder, rendering resonator arrays  a good backbone for the study of condensed matter physics with photons. 
\end{abstract}

\maketitle
Solving seemingly simple problems in quantum mechanics is a formidable task for even the most sophisticated classical computers. For this reason Feynman proposed using controlled quantum systems to simulate and study other quantum systems \cite{Feynman1982}. 
The development of such quantum simulators \cite{Buluta2009} has since been an active area of research in a number of physical systems, including ultracold atoms in traps and optical lattices \cite{Lewenstein2006, Bloch2008}, trapped ions \cite{Gerritsma2010}, and Josephson junction arrays \cite{Fazio2001}.

An idea of growing  interest is to use photons as particles in a quantum simulator for non-equilibrium systems \cite{Koch2009, Hartmann2006, Greentree2006, Angelakis2007, Hayward2012, Nissen2012,Houck2012}.   According to this idea, a photon lattice is created with an array of cavity quantum electrodynamics (cQED) elements, each consisting of a photonic cavity coupled strongly to a two level system, or qubit. Systems consisting of up to three coupled cavities have been realized for quantum information processing \cite{Johnson2010, Mariantoni2008, Wang2011}, and early proposals consider using larger arrays as a possible quantum computing architecture \cite{Helmer2009,DiVincenzo2009,Steffen2012}, but a lattice-based quantum simulator of this type has yet to be realized.

In these lattices, photons can hop between neighboring cavities and experience an effective photon-photon interaction within each cavity, mediated by the qubit.  The superconducting circuit architecture is an attractive candidate for realizing such lattices due to the flexibility afforded by lithographic fabrication and the relative ease of attaining strong coupling \cite{Blais2004}.  Such cQED lattices have been predicted to exhibit a wide variety of phenomena, including a superfluid-Mott insulator transition \cite{Koch2009, Hartmann2006, Greentree2006, Angelakis2007}, macroscopic quantum self trapping \cite{Schmidt2010}, and fractional quantum Hall physics \cite{Hayward2012}.

In order to jump-start the implementation of these lattice-based simulators, we have fabricated and characterized 25 arrays of cavities, with each cavity designed to be identical.(Devices discussed here do not include qubits yet.) In this letter, we focus on understanding and reducing uncontrolled disorder in arrays of resonators in a kagome geometry, the most natural two-dimensional lattice for such transmission line resonators.  We find that disorder in the individual resonator frequencies mainly originates from variations in the kinetic inductance due to small changes in the transverse dimensions of each resonator. We reduce disorder to less than two parts in $10^4$ with a suitable choice in the geometric layout of our transmission line resonators.

\begin{figure}
\centering
\includegraphics[width=8.7cm]{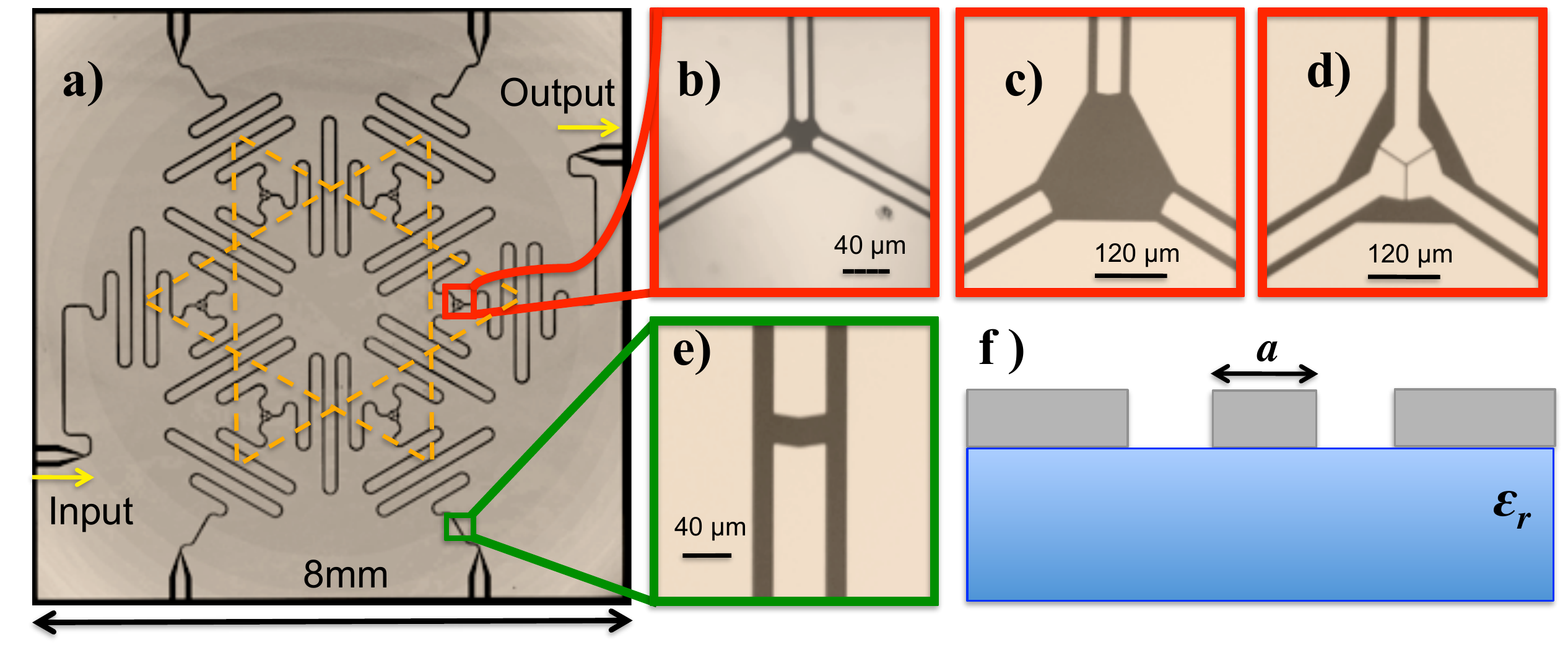}
\caption{{\bf{a)}} (Color online) Device picture of twelve capacitively coupled resonators. The overlaid orange dashed lines have been drawn between the coupled resonators and illustrate how the photonic lattice sites form a single kagome star. Transmission was measured between the ports labeled "Input" and "Output".{\bf{b,c)}} Images of symmetric 3-way capacitors with low hopping rate ($t/2\pi=0.8\,$MHz)  with 10$\,\mu$m  and 40$\,\mu$m wide center pins. {\bf{d)}}  Capacitor with high hopping rate ($t/2\pi=31\,$MHz) and 40$\,\mu$m wide center pin. {\bf{e)}} Image of outer coupling capacitor ($\kappa/2\pi=0.05\,$MHz) for 40$\,\mu$m center pin. {\bf{f)}} Cross-section of coplanar waveguide resonator with center pin width $a$, on a dielectric substrate $\varepsilon_r$.}
\label{DevicePic}
\end{figure}

The system of coupled cavities is described by the Hamiltonian:
%Hamiltonian equation
\begin{equation}
\label{Hamiltonian}
H=\sum_i \hbar(\omega_r+\delta_i)(a^{\dagger}_i a_i+\tfrac{1}{2})+\sum_{j>i} \hbar t_{ij}(a^{\dagger}_{j}a_i+a^{\dagger}_{i}a_{j})
\end{equation}
where  $a^\dagger_i, a_i$ are the photon creation and annihilation operators corresponding to resonator $i$ in the array.  We denote the frequency of resonator $i$ and its shift due to random disorder by $\omega_r$ and  $\delta_i$, respectively.
The hopping rate between resonators $i$ and $j$ is given by 
\begin{equation}\label{hopping}
t_{ij}=2Z_0C_{ij}(\omega_r+\delta_i)(\omega_r+\delta_j),
\end{equation}
where  $C_{ij}$ denotes the coupling capacitance between the cavities and $Z_0$ the characteristic impedance of the transmission line \cite{NunnenKamp2011}.
The array we study consists of twelve resonators coupled capacitively at their endpoints in a two-dimensional kagome geometry (Fig.~\ref{DevicePic}).  
Photon hopping was achieved by coupling triples of resonators in the interior of the array with  symmetric three-way capacitors. This coupling scheme naturally results in a kagome lattice. 

Uncontrolled disorder is a serious impediment to using such arrays for quantum simulation. Here, we work to understand the origins and minimize the level of disorder to mitigate this problem. 
In an ideal array with uniform resonator frequencies and hopping rate $t$, the spacing between normal mode frequencies  scales linearly with the photon hopping rate, with a frequency separation between the highest and lowest modes of $(3+\sqrt{5})t$; which arises from diagonalizing the 12x12 matrix resulting from  Eq.\ \eqref{Hamiltonian}.
Assuming that disorder in coupling capacitances is negligible, we find that disorder in resonator frequencies leads to shifts of the normal mode frequencies through the first term in Eq.\ \eqref{Hamiltonian} by an amount $\sim\delta_i$, whereas the additional disorder in $t_{ij}$ only results in corrections $\sim \delta_i t/\omega_r$. 
Since $\omega_r\gg t_{ij}$,  we can thus approximate the photon hopping rate to be uniform with value $t=2Z_0C\,\omega_r^2$ for nearest-neighbor resonators, and $t_{ij}=0$ for other resonator pairs. Therefore, we are primarily concerned with effects of disorder in resonator frequencies.

\begin{figure}
\centering
\includegraphics[width=8.8cm]{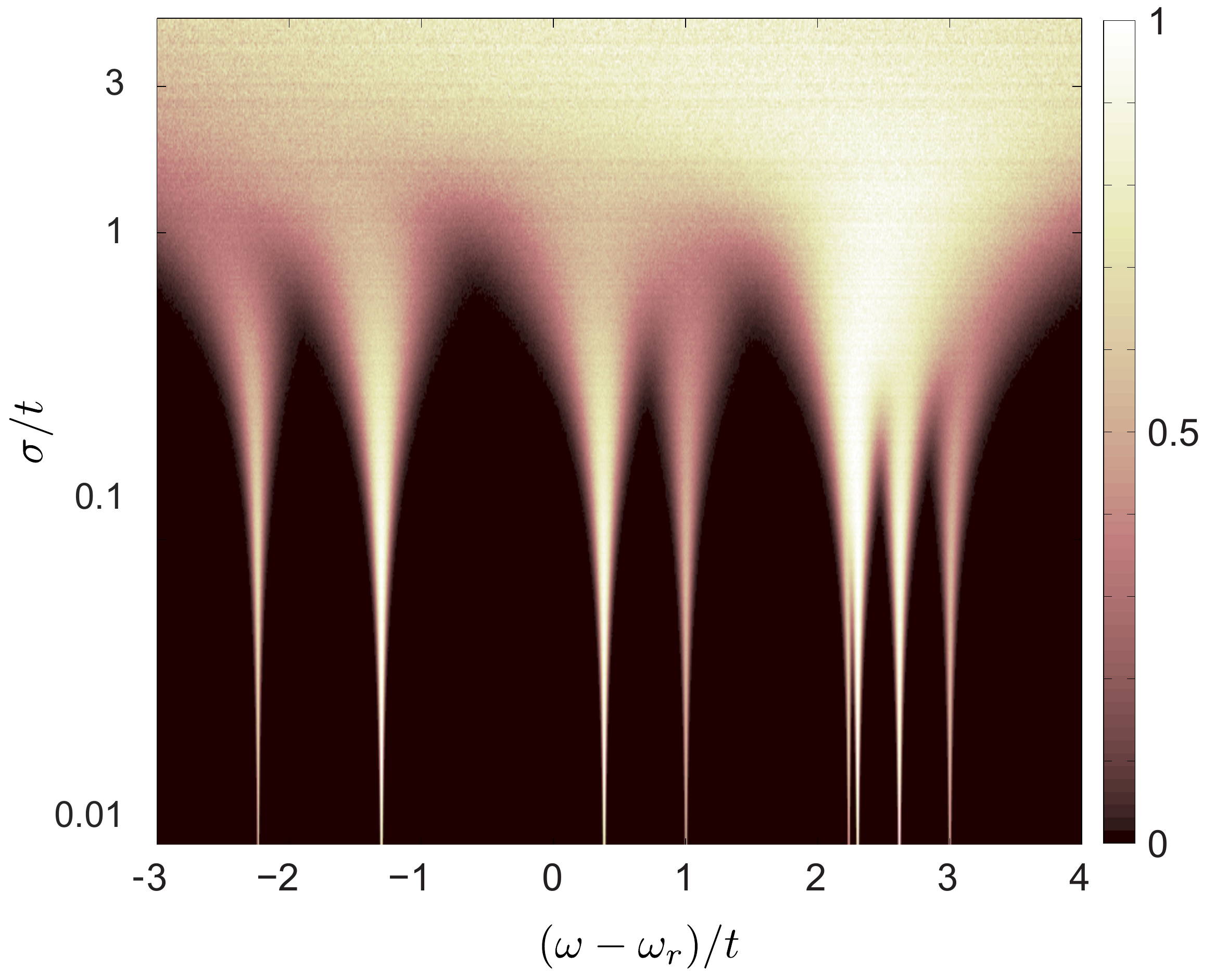}
\caption{(Color online) The normal mode histogram in the presence of disorder. 
Normal mode frequencies are calculated from Eq. \eqref{Hamiltonian} using a set of \{$\delta_i$\} drawn from a Gaussian distribution with standard deviation $\sigma$.  For each value of $\sigma$, this procedure is repeated $10^7$ times.  Histograms are generated from $10^7$ disorder realizations (for each value of $\sigma$), and are normalized to the maximum number of counts for clarity.   For $\sigma\gg t$, the histogram is dominated by disorder and forms a single Gaussian.  For $\sigma\ll t$, the histogram shows sharp peaks corresponding to the ideal normal mode frequencies. }
\label{PhotonDist}
\end{figure}

Without disorder, there are eight distinct mode frequencies, four of which are doubly degenerate. The presence of disorder breaks the degeneracy, widens the distribution of normal mode frequencies, and results in twelve distinct frequencies.  We study the effects of disorder by numerically diagonalizing the Hamiltonian for random \{$\delta_i$\} drawn from a Gaussian distribution with a standard deviation $\sigma$. The resulting histogram for the number of eigenmodes $N(\omega)\,d\omega$ in a given frequency interval $[\omega,\omega+d\omega]$ is shown in Fig.\ \ref{PhotonDist} for varying amounts of disorder $\sigma$. When $\sigma\ll t$, disorder is negligible and the normal mode frequencies are all close to those of the ideal lattice.  As $\sigma$ increases and becomes larger than $t$, the peaks in the distribution associated with individual normal mode frequencies broaden and ultimately merge.
Once merging occurs, the observed mode frequencies and corresponding modes can no longer be easily identified with the idealized modes. In the limit of $\sigma\gg t$, the normal mode histogram approaches a single Gaussian of width $\sigma$ from which the overall disorder of individual resonator frequencies can be extracted. For this reason, devices with a small hopping rate $t$ are ideal for discerning the effects of disorder.

%Plot of photon probability distribution

We have fabricated and measured $25$ arrays of twelve cavities to quantify disorder and assess the feasibility of quantum simulation in cQED lattices.   By design, each coplanar waveguide resonator had a frequency of $\omega_r/2\pi \approx 7\,$GHz, and an impedance Z$_0 = 50\  \Omega$. The devices were fabricated using photolithography on $200\,$nm of Nb sputtered onto a $500\,\mu$m thick sapphire substrate. Each device was mounted using high performance silver paste and then wire bonded to a copper circuit board. Wire bonds were also used to connect all ground planes. At the outer edges of the array, each cavity is capacitively coupled to a transmission line, resulting in a photon escape rate $\kappa=4 Z_0^2C_\text{out}^2 \omega_r^3$  to the continuum. This allows us to measure transmission  through opposite ports (Fig.~\ \ref{DevicePic}) of the array using a vector network analyzer. The unused ports were connected to 50$\,\Omega$ terminators, though no significant difference was observed when the ports were left open.  Each device was cooled to a base temperature of $20\,$mK inside a dilution refrigerator -- a necessary requirement for future quantum simulations with small numbers of polaritons \cite{Koch2009, Hartmann2006, Greentree2006, Angelakis2007, Hayward2012, Koch2010}.

The set of our 25 devices, summarized in Table \ref{Results}, includes samples with two distinct hopping rates of $t/2\pi=0.8\,$MHz and $t/2\pi=31\,$MHz. These nominal values were obtained from Eq.\ \eqref{hopping} by using values for the coupling capacitences determined using a finite-element calculation. While the high-$t$ devices allow us to access $t\gg\sigma$ and are most useful for quantum simulation, the low-$t$ devices are the better choice for characterizing disorder. 

%Table of results
\begin{table}
\caption{\label{Results} Results extracted from 25 measured devices.  Devices were characterized with two different photon hopping rates $t$ and three different center pin widths $a$. The random disorder $\sigma$ was extracted from peak positions of the transmission spectrum for each device.  The disorder decreases for increasing $a$. The ratio $\sigma/t$ is a metric of how the normal mode frequencies are effected by disorder. For the $40\mu$m devices, $\sigma$ is reduced to less that two parts in $10^4$ of $\omega_r/2\pi$. All uncertainties are computed from standard deviation of individual measurements. }
\begin{ruledtabular}
\begin{tabular}{c c c c c}
$t/2\pi$ (MHz) & $a$ ($\mu$m) & $\sigma/2\pi$(MHz) & \quad {\bf{$\sigma/t$}} & \quad $\#$ Measured\\
\hline
0.8 & 10 & $9.1\pm 2.8$ &\quad 11.5 & \quad $13$\\
0.8 & 20 & $3.9\pm 1.2$ &\quad 4.9 & \quad $4$ \\
0.8 & 40 & $1.4\pm 0.8$ &\quad 1.7 & \quad $4$\\ 
31 & 40 & $1.3\pm 0.3$&\quad 0.04 & \quad $4$\\ 
\end{tabular}
\end{ruledtabular}
\end{table}

\begin{figure}
\centering
\includegraphics[width=8.7cm]{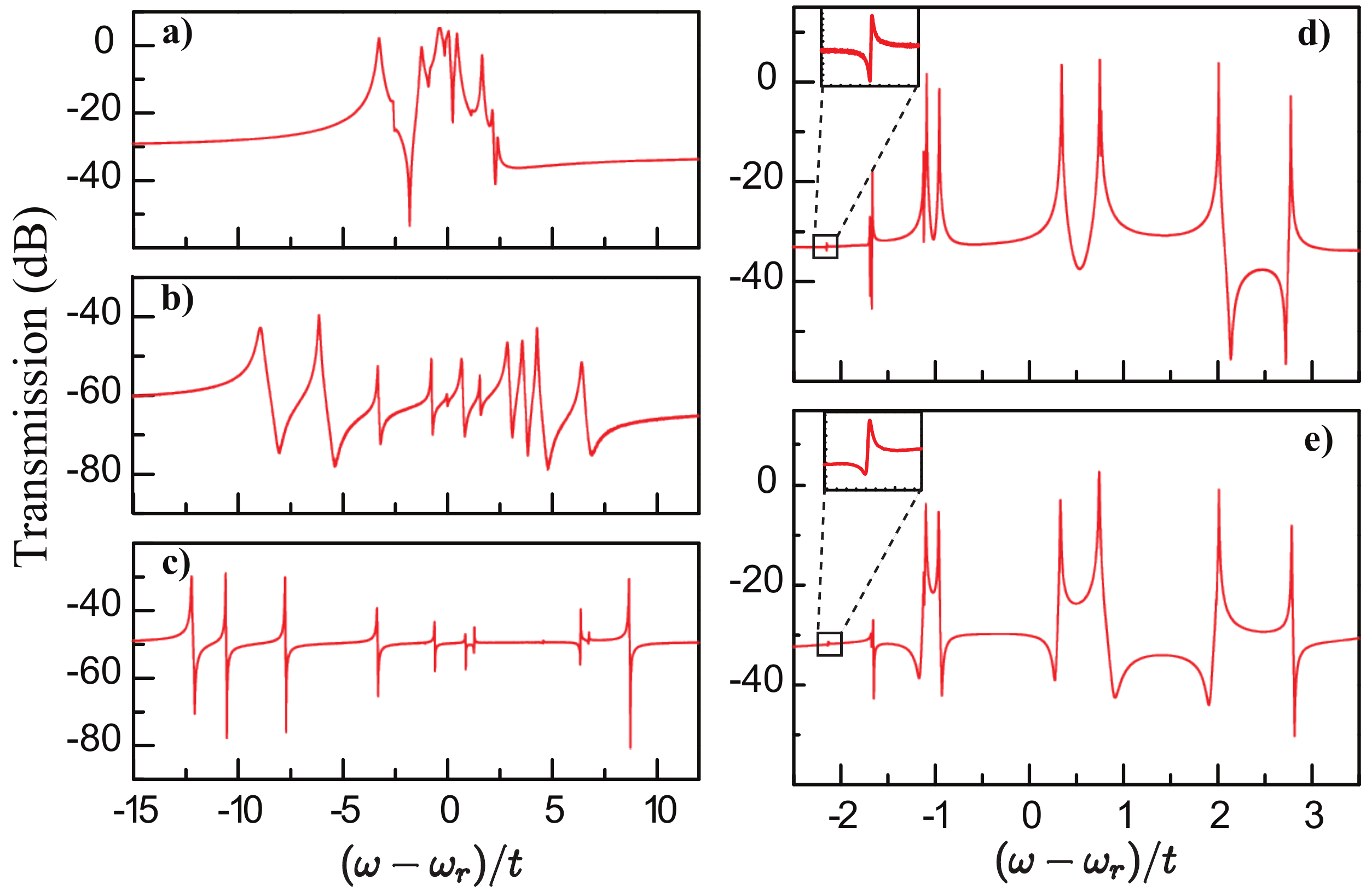}
\caption{Transmission spectra for measured devices.  The first column shows spectra for devices with {\bf{a)}}  $t/2\pi = 0.8\,$MHz, $a=40\,\mu$m,  {\bf{b)}} $t/2\pi = 0.8\,$MHz, $a=20 \,\mu$m, {\bf{c)}} and $t/2\pi = 0.8\,$MHz, $a=10\,\mu$m. The width of the spectrum  decreases for increasing resonator width, demonstrating a decrease in $\sigma$.  The second column {\bf{d),e)}} shows transmission spectra for two nominally identical devices with $t/2\pi = 31\,$ MHz and $a= 40 \,\mu$m. Each scan contains twelve well defined peaks that are consistent between the two devices. Peak positions are similar to those expected, when accounting for a systematic edge effect due to the difference between inner and outer capacitors. The inset shows the lowest energy mode that is  localized on the inner six resonators in the absence of disorder.}
\label{Scans}
\end{figure}

We extract  normal mode frequencies from the peak positions in the measured transmission spectra (Fig. \ref{Scans} {\bf{a)-c)}}) in order to determine the disorder. To account for small systematic shifts in devices made in separate fabrication batches, all frequencies were expressed relative to the mean peak frequency of each spectrum. For  low-$t$ devices, not all twelve peaks are always visible. Such ``missing" peaks can be due to normal mode degeneracies (occuring in the ideal case), as well as normal modes with small or vanishing amplitude in either of the resonators coupled to the input or output port.

% Transmission Spectra plot

For low-$t$ devices, analyzing the peak positions provides a systematic method for extracting $\sigma$ from a transmission measurement.
Specifically, the disorder strength can be extracted from the peak positions using:
\begin{align}
\label{RandomDisorderEqn}\nonumber
\sigma^2&=\left\langle \frac{1}{n}\sum_{i=1}^{n} \delta_i^2\right\rangle\\
&=\left\langle \frac{1}{n}\sum_{i=1}^{n}\left(\Omega^\text{dis}_{i}-\bar\Omega^\text{dis}_{i}\right)^2 \right\rangle
-\frac{1}{n}\sum_{i=1}^{n}\left(\Omega_{i}-\bar\Omega_i\right)^2,
\end{align}
where $n=12$ is the number of resonators in each sample, $\Omega_i$ and $\Omega_i^\text{dis}$ denote the twelve normal mode frequencies in the absence and presence of disorder, respectively. 
$\bar\Omega_i$ and $\bar\Omega_i^\text{dis}$ are their means (for a single disorder realization),  whereas ensemble averages over disorder realization are denoted by $\langle \cdot \rangle$. In the disorderless case, the ``variance"  of the normal mode frequencies of the kagome star is $3t^2$.

Applying this method to samples with a standard 10$\,\mu$m width of the transmission line center pin, we find that the disorder $\sigma/2\pi=(9.1\pm2.8)\,$MHz is larger than expected from resonator length variations due to finite resolution in optical lithography. To investigate the origin of this disorder, we fabricate devices with different widths $a$ of the center pin, while maintaining a constant $Z_0$ throughout and find that there is a systematic dependence of disorder on $a$.

\begin{figure}
\centering
\includegraphics[width=8.7cm]{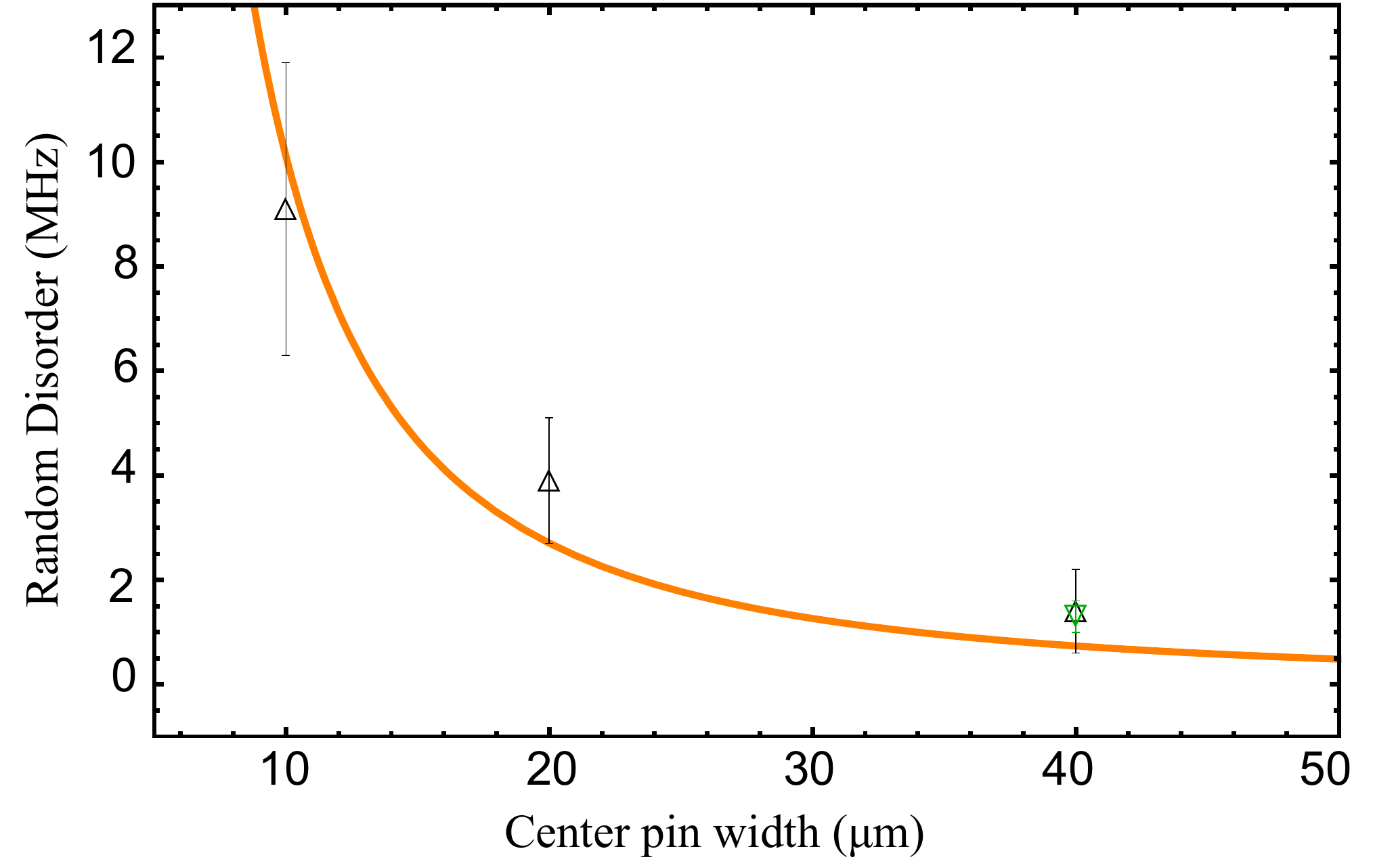}
\caption{ Random disorder versus center pin width for all devices. Disorder extracted from low-$t$ devices is plotted in black with upward pointing triangles, while disorder extracted from high-$t$ devices is plotted in green with a downward pointing triangle.  The curve shows the difference in frequency for two resonators, one with center pin width equal to the value on the horizontal axis and the other with a center pin width $600\,$nm smaller and dielectric gap $1200\,$nm larger. Error bars are computed from standard deviation of individual measurements.}
\label{RandomDisorder}
\end{figure}

The magnitude of disorder decreases with increasing center pin width (Fig.~\ref{RandomDisorder}).  This dependence of disorder on the device geometry can be attributed to random variations in the width of the center pin that arise during microfabrication. These variations in width result in variations in the kinetic inductance $L_k$, which in turn affects the resonator frequency through the relation: 
\begin{equation}
\label{ResFreq}
\omega_r=\frac{1}{2\sqrt{(L_{m}+L_{k}) C_{\text{tot}}}}
\end{equation}
where $L_m$ is the intrinsic magnetic inductance and $C_{\text{tot}}$ is the total capacitance. In normal metals, $L_k$ is suppressed by electron scattering but in superconductors the DC electrical resistance is vanishing and $L_k$ is no longer suppressed. Although $L_k$ is more relevant in superconductors, it is still two orders of magnitude smaller than $L_m$, for the device geometry considered here. 
For a single resonator, $L_k$ typically only results in a small shift in $\omega_r$ \cite{Frunzio2005,Fragner2008}. For arrays of coupled resonators, however, these small shifts can introduce significant disorder if the kinetic inductance contributions vary across the array. 

For the small length scales used here, the sensitivity of the kinetic inductance to variations in $a$ decreases rapidly as the width $a$ is increased \cite{Watanabe1994}\footnote{The expression for $L_k$ from \cite{Watanabe1994} is accurate for film thickness less than twice the penetration depth. It was used here to obtain a rough estimate for the geometric dependence of the superconductor}.  
 
 In our devices, we observe variations in the center pin width of up to $\sim 600\,$nm and twice that for the dielectric gap, when examining them with a scanning electron microscope.  The random disorder expected due to kinetic inductance variations can be estimated by comparing $\omega_r$ for cavities of equal length but with widths differing by the observed $600\,$nm, see Fig.~\ref{RandomDisorder}. The random disorder observed here is consistent with variations in device geometry and can be reduced to less than two parts in $10^4$ by making resonators with $40 \mu$m wide center pins. 

Using this strategy to reproducibly obtain devices with small disorder, we next turn to the high-$t$ devices. Transmission spectra for all four of these devices revealed very similar normal mode frequencies, confirming that disorder was small. Two representative transmission spectra are shown in Fig.~\ref{Scans} {\bf{d),e)}}. For all high-$t$ devices, the lowest energy mode is significantly smaller in amplitude than the other eleven modes. In the absence of disorder, the lowest energy mode is localized to the six inner resonators and cannot be driven from any port. For the infinite kagome lattice, it is these localized states that lead to the known flat bands \cite{Koch2010,Mielke1991,Mielke1992,Mielke1993}. Disorder in the array weakly breaks this localization and causes the mode to acquire a small amplitude in the outer resonators.

For high-$t$ devices, $\sigma$ is small compared to $t$ and both variances on the right-hand side of Eq.\ \eqref{RandomDisorderEqn} are large and nearly cancel each other. Consequently, an alternate method to extract $\sigma$ in these devices is used.  In devices where $t\gg \sigma$, the twelve peaks are easily identifiable in the transmission spectra and directly indicate the variation in individual normal mode frequencies. In this limit, the frequency $\Omega_j^{\text{dis}}$ of the $j^{th}$ normal mode can be expanded to lowest order in the \{$\delta_i$\} as
\begin{equation}
\Omega^\text{dis}_j=\Omega_j+\sum_i \frac{\partial \Omega_j}{\partial
\delta_i} \delta_i.
\end{equation}
The variance of this normal mode frequency with respect to disorder is then
\begin{align}
\label{highTdisorder}\nonumber
\sigma_j^2&=\left<\left(\Omega_j^\text{dis}\right)^2\right>-\left<\Omega_j^\text{dis}\right>^2\\
&=\sum_i \left(\frac{\partial \Omega_j}{\partial \delta_i}\right)^2
\sigma^2.
\end{align}
The partial derivatives in Eq.\ \eqref{highTdisorder} can be calculated numerically.  Doing so, one finds that for each pair of degenerate normal modes the two normal mode frequencies depend on mutually exclusive sets of the \{$\delta_i$\}.  Thus, these two eigenfrequencies fluctuate independently about a common mean value. In order to estimate disorder $\sigma$ from the measurements of the high-$t$ devices, we first calculate the variance of the frequencies corresponding to each set of singly or doubly degenerate normal modes. Then, using Eq.\ \eqref{highTdisorder}, we scale this variance by the sum of the squares of the partial derivatives to calculate an estimate for the disorder $\sigma$.  Finally, we take the average of the estimates for $\sigma$ found for all of the sets of normal modes, weighted by the order of the degeneracy of each set. Using this method, we find $\sigma=(1.1\pm0.6)\,$MHz which is well into the limit of $\sigma\ll t$.

The measured peak positions for the high-$t$ devices agree well with expectations when accounting for a systematic shift between inner and outer resonators due to the difference between inner and outer capacitances.  In principle, the systematic shift can be compensated by adjusting the length of the edge resonators and will be of less importance in large arrays. As disorder is now understood, building larger arrays and adding qubits are the next logical and feasible steps.

Photon lattices open the door for future experiments looking for quantum phase transitions and other many-body photon effects in coupled cQED arrays. Uncontrolled disorder in arrays of bare cavities can be reduced to only a few parts in $10^4$. With random disorder sufficiently reduced, cavity arrays can be fabricated with controlled amounts of disorder in order to study localization effects. \cite{Anderson1956,Schwartz2007}.   While fabricating qubits on resonance with each cavity will be challenging, strong effective photon-photon interactions have already been observed even when qubits are off resonance \cite{Hoffman2011}.  This strong off-resonant interaction should enable the possibility of studying many-body effects in the presence of disorder in qubit frequency, as resonant cQED is not a strict requirement.  Therefore, quantum simulation with cQED lattices appears to be a realizable goal with superconducting circuits.
 
\begin{acknowledgments}
We thank S.\ M.\ Girvin
for valuable discussions. This work was supported by the National Science Foundation through grant nos.\  DMR-0953475%(Princeton)
, PHY-1055993%(JK)
, by the Army Research Office under contract  W911NF-11-1-0086%(Princeton)
, and by the Packard Foundation. Additionally, D. Underwood is supported by a National Science Foundation Graduate Research Fellowship DGE-1148900.
\end{acknowledgments}

\bibliography{KagomeCitations}

\end{document}